\documentclass[manuscript,screen]{acmart}
\usepackage{enumitem}
\usepackage[sorting=none]{biblatex}
\usepackage{xcolor}
\usepackage{lipsum}

\newcommand\blfootnote[1]{%
  \begingroup
  \begin{NoHyper}
  \renewcommand\thefootnote{}\footnote{#1}%
  \addtocounter{footnote}{-1}%
  \end{NoHyper}
  \endgroup
}

\usepackage{hyperref}
\hypersetup{
    pdftitle={Unlocking Innate Computing Abilities in Electric Grids},    
    pdfauthor={},    
    pdfsubject={},   
    pdfcreator={},   
    pdfproducer={},  
    pdfkeywords={},  
    colorlinks=false,         
    linkcolor=blue,           
}

\begin{document}

\title{Unlocking Innate Computing Abilities in Electric Grids}

\author{Yubo Song}
\email{yuboso@energy.aau.dk}
\orcid{0000-0003-1452-449X}
\author{Subham Sahoo}
\email{sssa@energy.aau.dk}
\orcid{0000-0002-7916-028X}
\affiliation{%
  \institution{Department of Energy, Aalborg University}
  \city{Aalborg}
  \country{Denmark}
}

\blfootnote{
e-mails: \texttt{\{yuboso, sssa\}}@energy.aau.dk
}

\begin{abstract}
Electric power grids are engineered energy systems whose forward electrical responses embody high-dimensional and memory-bearing transformations of input signals. In this work, we reveal that these transformations---inherent in electric circuit elements, power flows and network topologies---can be conveniently harnessed for computation without modifying physical grid architectures. By encoding structured input data into the operational setpoints of power electronic converters inside grids, we demonstrate how forward grid dynamics are interpreted into physical representations comprising system variables---by showcasing through an affine transformation example implemented on a direct-current (DC) grid---which justifies the capability of grids performing information processing tasks concurrently alongside normal power flows. Our work not only underscores the computation capability intrinsic to grid physics, but also opens a new perspective on how energy networks can function as sustainable computational substrate. This positions them as flexible assets where several computing tasks from data centers can be sustainably outsourced.
\end{abstract}

\renewcommand\keywordsname{Key Words}
\keywords{Smart Energy Systems | Electric Grids | Physical Computing | System Network Responses | Power Electronics | Sustainable Infrastructures}

\maketitle


\section*{Introduction}

Over the years, digital computations in data centers have pushed electric power grids to their tipping limits due to insufficient energy availability. Albeit the advances in algorithms, specialized hardware (processors) and edge-computing architectures have improved the energy efficiency, the aggregated energy consumptions associated with computation in scaled engineered systems remain spanning in the order of terawatt-hours (TWhs) annually from electric grid infrastructure \textcolor{blue}{\cite{energyai}}. This fact has escalated to a point where challenges emerge in electric grid operation, energy storage management and de-carbonization trajectories \textcolor{blue}{\cite{jevonparadox}}---highlighting the need of new physical computations in the environment that enable learning using existing physical signals, such as electricity, light, and material properties \textcolor{blue}{\cite{annualreviews}}.

From a physical standpoint, computation is not tied to digital algorithms and \textit{in silico} infrastructures. At its core, computation amounts to structured transformation of input signals by physical systems into reproducible responses through deterministic dynamics---including both instantaneous gains and spatial/temporal signal propagation---subject to physical laws \textcolor{blue}{\cite{jaeger2023computingphys}}. This underlies a broad class of computing paradigms, ranging from biological neural systems that inspire neuromorphic schemes \textcolor{blue}{\cite{elishai2022NeuroEng}} and artificial dynamical systems that form the basis of physical reservoir computing networks \textcolor{blue}{\cite{yan2024reservoir, benjamin2025phxresistive}}. Exploiting such mechanisms offers a promising pathway towards computation with reduced \textit{energy overhead} from specialized digital architectures by aligning the signal transformation with the naturally occurring physical processes.

Electric power grids are one of the scaled and complex engineered dynamic systems, where its behavior is governed by electric circuit laws, power flow constraints and network topologies \textcolor{blue}{\cite{kundurpowersys, dirk2022revphx}}, giving rise to high-dimensional signal transformation capabilities both spatially and temporally. Perturbations introduced into the network---episodic changes in power generations, loads or control references---will propagate through the physical network and control paths, producing predictable system responses and naturally exhibiting the same intuition as digital processors.

This is further underscored in modern electric grids, particularly dominated by power electronics, where:
\begin{enumerate}[nosep, label=(\arabic*)]
    \item Power electronic converters interfacing renewable energy sources and loads introduce exceptional degrees of programmability through feedback control,
    \item Distributed network and electrical states naturally give rise to high-dimensional (both linear and non-linear) responses,
    \item Temporal coupling introduce intrinsic memory (historic states) to the system dynamics,
\end{enumerate}
that justifies their potential in computational ability \textcolor{blue}{\cite{jake2023phxspatial}}. By regulating the grid behaviors in response to input signals, the controllers are essentially shaping the associated plant transfer functions, which are equivalent to network impedances \textcolor{blue}{\cite{tcgreen2021circuitmodel}}. As a result, the propagation of power flow across the network is handled using signal interfaces at converter terminals. Importantly, such regulation is typically subject to standard operational constraints of the physical system and co-exist with the primary objective of designated power transmission.

Despite its dynamic richness and large controllability, electric grids have solely been studied from a monolithic perspective of performance metrics including efficiency, stability and reliability, etc \textcolor{blue}{\cite{entsoroadmap}}. Hence, the idea of opportunistically leveraging their forward dynamics in terms of information-processing framework remains largely unexplored till date. In this work, we therefore establish a novel framework that embodies this concept and exploit electric grids as powerful computing asset. By encoding structured input data into the reference signals of controllable components, i.e., power electronic converters, we reveal that the resulting system responses constitute physical representations of the target computed outputs. The corresponding input-output mapping is determined by the network topology, electrical parameters and control mechanisms, allowing it to be systematically designed for reproducible computational tasks.

As a concrete demonstration, we exemplify an elementary affine transformation task for images using a five-converter direct-current (DC) microgrid. In this system, we utilize droop control \textcolor{blue}{\cite{dcdroop}} to realize tunable impedances that directly correspond to adjustable signal propagation weights, of which the underlying rationale is extensively elaborated. The computation emerges intrinsically from the electrical dynamics and occurs concurrently alongside normal power flow, without requiring additional hardware or compromising the rated operating conditions. While validated on a small-scale microgrid architecture, the underlying physical rationale is promisingly applicable and scalable to embedding computation within general engineered dynamical systems.

\section*{Electric Grids as Physical Computing Substrate}

\begin{figure}[H]
    \centering
    \includegraphics[width=0.6\linewidth]{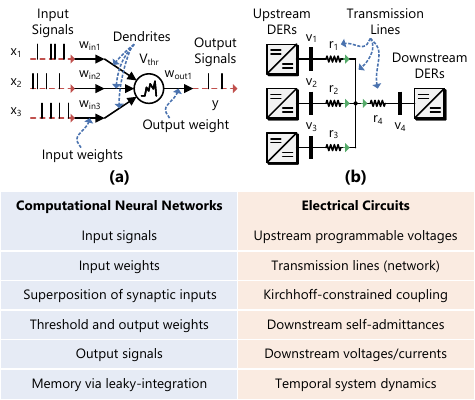}
    \caption{Conceptual positioning of electric grids as physical computing substrates, where (a) a spiking neural network exemplifies computation emerging from weighted superpositions, temporal integration and nonlinear activations. This occurs similarly in (b) electric grids, where interconnections through transmission lines and control flexibility owing to power electronic converters intrinsically enable the physical system states (voltages, currents, power flows and corresponding reference inputs) to be transformed through fundamental circuit laws and network dynamics.
    }
    \label{fig_comparison}
\end{figure}

The conceptual positioning of electric grids with regard to computing can be understood by examining how computational tasks are performed. A representative example, the Spiking Neural Network (SNN), which is inspired by the circuit-like dynamical physics of biological neurons and studied as an efficient data transformation approach, is depicted in \textcolor{blue}{Fig.~\ref{fig_comparison}(a)} \textcolor{blue}{\cite{roy2019neuromorphic, jason2023snntraining}}. In SNNs, computation arises from the collective interaction of weighted couplings among neurons, temporal integration of neural responses, and linear/nonlinear functions including activation/thresholds.

Remarkably, modern electric grids, comprising numerous distributed energy resources (DERs) interfaced through power electronic converters (PECs), naturally exhibit close resemblance to such input-output transformation behaviors \textcolor{blue}{\cite{spiketalk}}. As shown in \textcolor{blue}{Fig.~\ref{fig_comparison}(b)}, power flows among upstream \& downstream DERs are governed by intrinsic steady-state and dynamical (differential) circuit equations---where Ohm's law dictates the scaling between voltages and currents in proportion to network impedances, and Kirchhoff's current law enforces the aggregation of the currents \textcolor{blue}{\cite{circuitbook}} (as translated from weighted voltages) at terminal nodes or intermediate points of common coupling (PCCs). As such, the externally imposed control references can be mapped into measurable physical states through the grid's forward electrical dynamics. This structural analogy, rooted in comparable principles between power flow and information propagation, opens up transformative opportunities for leveraging electric grids as physical substrates for computation.

This conceptual alignment is further illustrated in \textcolor{blue}{Fig.~\ref{fig_framework}}, in which we present an affine transformation example (detail elaborated in \textcolor{blue}{Methods}), that is often used as an elementary scenario for neural-network mapping from a mathematical perspective. At its core, an affine transformation is defined as the linear operation:
\begin{equation}
    \mathrm{\mathbf{y} = \mathbf{Wx}+\mathbf{b}}
    \label{eq_affine}
\end{equation}
where, $\mathrm{\mathbf{W}}$ and $\mathrm{\mathbf{b}}$ denote the weight matrix and bias vector, respectively. In neural networks, such linear mapping is typically followed by nonlinear activation or thresholding operations that enable more expressive input-output relationships.

A close transformation exists in electric grids as described by the network-level Ohm's Law:
\begin{equation}
    \mathrm{\mathbf{I} = \mathbf{GV} = \mathbf{G(V^*+\Delta V)}}
    \label{eq_grid}
\end{equation}
where, $\mathrm{\mathbf{G}}$ is the network conductance matrix (or admittance in non-resistive cases), $\mathrm{\mathbf{V^*}}$ and $\mathrm{\mathbf{\Delta V}}$ represent the nominal operational points and its controller-induced deviation, respectively. The key observation is that the controllability of $\mathrm{\mathbf{V^*}}$ and $\mathrm{\mathbf{\Delta V}}$ facilitates a direct correspondence between the mathematical affine transformation in \textcolor{blue}{(\ref{eq_affine})} and the physical electrical transformation in \textcolor{blue}{(\ref{eq_grid})}. In this context, the affine weights $\mathrm{w_{kl} = {\partial y_k}/{\partial x_l}}$ can directly be implemented with the network conductances $\mathrm{g_{kl} = {\partial i_k}/{\partial v_l}}$, grounding the computing ability into the intrinsic physics of electric grids. More importantly, this mapping can be realized through the grids' forward electrical response and does not require modifications of the core physical grid architecture.

\section*{Results}

Using these theories, elementary cases are demonstrated based on \textcolor{blue}{Fig.~\ref{fig_framework}(b)}, where $2 \times 2$ pixel images are processed as input data through the grid as per affine transformation, with its weights $\mathrm{w_{k}}$ specified in \textcolor{blue}{Methods}. The current variations $\mathrm{\Delta i_5}$ in each case are measured as output data and decoded to associated images.

The heatmaps in \textcolor{blue}{Fig.~\ref{fig_linresults}}\textcolor{blue}{(a1)} and \textcolor{blue}{(b1)} visualize the current variations $\mathrm{\Delta i_k}$ at DER $\mathrm{k}$ in response to the input data posed as $\mathrm{\Delta V_{ref,k}}$ that are selectively applied to the DERs. For the case study in \textcolor{blue}{Fig.~\ref{fig_linresults}(a1)}, the resulting current variations $\mathrm{\Delta i_5}$---subject to the equivalent conductances $\mathrm{g_{eq, 5k} = {\partial i_5}/{\partial V_{ref,k}}}$ and steady-state electric circuit laws---align proportionally with the predefined computational weights $\mathrm{w_{k}}$ (corresponding to a quantization level of -0.09 A). While the transmission system parameters in $\mathrm{\mathbf{G}}$ remain untouched in the case study in \textcolor{blue}{Fig.~\ref{fig_linresults}(b1)}, adapting the droop gains of each DER enables the reshaping of the equivalent weight matrix $\mathrm{\mathbf{W}}$. Hence, this provides empirical evidence supporting the feasibility of utilizing electric grids to perform computations by simply programming the control layer of each DERs---without any infrastructural changes in transmission lines and DERs themselves.

In \textcolor{blue}{Fig.~\ref{fig_linresults}}\textcolor{blue}{(a2)} and \textcolor{blue}{(b2)}, we extend the analysis from one-hot inputs to simultaneous activation of multiple pixels (interlaced inputs), corresponding to multiple nonzero $\mathrm{\Delta V_{ref,k}}$. The affine computing network and electric grid perform matrix operations within their underlying network using weighted summation of input data and physical conductance-governed power flow, respectively---the outputs in both cases adhere to the linear superposition principle. This behavior is clearly reflected in the aggregated current responses in \textcolor{blue}{Fig.~\ref{fig_linresults}}\textcolor{blue}{(a2)} and \textcolor{blue}{(b2)}, justifying its potential for innate distributed computation in electric grids across high-dimensional input space.

We perform time-domain demonstrations in \textcolor{blue}{Fig.~\ref{fig_waveforms}} with circuit-level simulations to physically substantiate the proposed framework from an application perspective. Starting from the \textit{static} operation point, step perturbations $\mathrm{\Delta V_{ref,k}}$ are applied to the voltage references as encoded computational inputs. The resulting currents closely match the numerical results in \textcolor{blue}{Fig.~\ref{fig_linresults}} that are subject to Kirchhoff-constrained network behaviors, and the system converges to new equilibrium states that are then decoded into computational outputs as per the aforementioned scheme. Notably, these waveforms therefore visualize the steady-state mappings used for computation and confirm that the ability is realized inherently through the grid's forward electrical responses. Meanwhile, the observed settling characteristics are governed by system dynamics rather than computational clocks, thus the ability should not be assessed solely through a speed-centric proposition. Instead, it introduces additional dimensions of computational states beyond steady-state mappings by leveraging the rich transient dynamics especially when the system scales up.

Moreover, the energy footprint of the proposed framework can be further evaluated based on \textcolor{blue}{Fig.~\ref{fig_waveforms}}. In the demonstrated cases, each DER is rated at 1 kW by default, and the baseline power loss on the lines totals 32.74 W. By comparison, the losses in the computation cases \textcolor{blue}{(a)} and \textcolor{blue}{(b)} reach to 91.38 W and 40.26 W, respectively---corresponding to additional power consumption equivalent to 1.17\% and 0.15\% of the total power generation. Across the inference process as a whole, this additional energy consumption, however, is determined by the magnitude and duty of the deviations $\mathrm{\Delta V_{\mathrm{ref},k}}$ relative to the nominal operating point, together with the statistics of the encoded signals, given that computation is encoded as voltage/current deviations superimposed on existing power flows, and not a continuous cycling or high-frequency excitation that stresses the components. In contrast to conventional digital computing where energy consumption is dominated by dedicated processing hardware, this energy consumption arises alongside the regular energy flow without additional infrastructural energy overhead, and can be further minimized by optimizing the encoding scheme according to the average deviation for typical input statistics---where the computing accuracy is not compromised and the component reliability can still be ensured well-contained and manageable. This justifies the framework's potential in energy footprint towards more sustainable computing substrates.

\section*{Discussion}

In this study, we explore the structural and functional similarities between computing networks and electrical grids, demonstrating how these parallels can unlock their inherent computational capabilities for the first time. By encoding and processing data through system operations, we reveal that the forward dynamics in electric grids support both energy transmission and information processing. Although our examples are simple, the scientific perspective re-frames electric grids as novel multifunctional physical substrate. More broadly, as the affine transformation demonstrated here is a fundamental building block of many computational workloads, it also opens new avenues for harnessing electric grids or more general engineered systems as powerful compute assets for linear transformations (e.g., convolution or finite-impulse-response (FIR) filtering), affine layers of artificial intelligences (AIs) or distributed optimization tasks (e.g., gradient or consensus computations), that data centers can leverage to virtually expand their computing capacity.

Stepping beyond our work, we also offer our perspectives and outline key challenges for future research:
\begin{enumerate}[label=(\arabic*)]
    \item While resistive networks with droop control are well suited for linear affine transformations, the framework can be generalized to alternating-current (AC) grids leveraging complex phasor and direct-quadrature-zero (dq0) axis decomposition, as well as energy conservation when accounting for DC/AC interconnections. Besides, we recognize the necessity of nonlinear activation functions to realize the full computational capabilities of more expressive multi-layer computational architectures including neural networks, which can potentially be integrated into the programmable controllers of DERs. Meanwhile, we encode a pixel with high intensity into a unit increase of $\mathrm{V_{ref,k}}$. Nevertheless, this framework can be extend to more complex dynamical regimes when assisted by forefront information-coding techniques, particularly in higher-dimensional alternating-current (AC) grids, where rich temporal \& spatial impedance characteristics and nonlinearities could naturally support more scaled computations to operate with increased information-processing capability.

    \item The proposed computing framework does not rely on any digital communication among DERs---all computations and information propagation are completely physical with no digital footprint. This eliminates the conventional entry points for cyber attacks, guaranteeing its resilience against any digital spoofing or manipulation methods. Besides, the impacts of noise or perturbations are also localized and do not accumulate along any communication path, indicating another inherent advantage over digitally cascaded signal chains.

    \item In terms of scalability, the affine mapping for larger networks is obtained directly by extending the known conductance matrix $\mathbf{G}$ to corresponding Jacobian $\partial\mathbf{i}/\partial\mathbf{V}_\mathrm{ref}$---as an offline design-time parameter assignment of $\mathrm{\Delta R_{\mathrm{d},k}}$ instead of real-time operational coordination among converters. This implies a bounded complexity $\mathrm{O(N^2)}$ for initialization of the framework where $\mathrm{N}$ stands for number of physical nodes, typically different from data-driven substrates.
    
    \item Training may emerge as a challenge in the context of electric-grid-based neural computation, as it also relies on control-layer programmability apart from traditional backpropagation algorithms. Similarly, the neural network \textit{weights} can be translated as control gains or grid-edge solutions, etc., while effective training necessitates not only a tailored cost function but also the compliance with grid operational constraints like system stability and power flow convergence.

    \item This computing mechanism shares common premises with physics-based and reservoir computing in that both exploit physical systems' intrinsic forward dynamics for computation. However, reservoir computing typically requires the substrates to be \textit{resettable} through equilibrium propagation across two operational states: free and nudged phases, and often presupposes the substrates to be purpose-built for this procedure. Such requirements are not readily compatible with electric power grids, which are large-scale, continuously operating and cannot be reset or driven through repeated free/nudged cycles---a constraint that our framework can override by shaping the system analytically from its known physics. Nevertheless, physics-based and reservoir computing retain advantages in robustness to uncertainties and task generality that inspire future improvements of our framework towards expanding the computing ability of grids as promising edge computing assets.
    
    \item Since the computation mechanisms are intrinsic to electric grid operation---emerging naturally from the physical laws that govern voltages, currents and power flows---as well as more general engineered systems, the proposed approach holds promising potential to reshape the energy footprint of conventional digital computational or information-processing systems through integrated infrastructures, and outsource AI workloads from centralized processors and data centers for the green transition.
\end{enumerate}

\newpage
\section*{METHODS} 

Here we provide a structural description of the study case of our article, as well as analytical derivations underlying the computing framework. Further details regarding the methods can be found in the supplemental information.

\subsection*{Case Description}

\subsubsection*{Studied Electric Grid}

The exemplary case of this article is a DC microgrid (\textcolor{blue}{Fig.~\ref{fig_framework}(b)}) consisting of 5 distributed energy resources (DERs) interfaced by DC-DC power electronic converters (PECs), with resistive power lines and resistive loads locally connected at each DER bus. Voltage regulation at each DER is implemented by digital feedback controllers designed to follow the DC droop control equation for decentralized proportional current sharing (\textcolor{blue}{Fig.~S1}) \cite{dcdroop}:
\begin{equation}
    \mathrm{V_k^* = V_{\mathrm{ref},k}+R_{\mathrm{d},k}i_k},
    \quad \mathrm{k} \in \{1,2,3,4,5\}
    \label{eq_droop_appdx}
\end{equation}
where, for each DER $\mathrm{k}$, the terminal voltage reference $\mathrm{V_k^*}$ (references for primary voltage regulators) is shifted from the rated reference $\mathrm{V_{\mathrm{ref},k}}$ by an amount proportional to the source current $\mathrm{i_k}$. The droop gains $\mathrm{R_{\mathrm{d},k}}$ enable proportional current sharing of the DERs, and equivalently act as virtual resistances from system perspective.

In the proposed framework, the innate computing ability in transmission network circuits is leveraged by manually shifting each droop gain $\mathrm{R_{\mathrm{d},k}}$ (which is technically programmable) with an offset $\mathrm{\Delta R_{\mathrm{d},k}}$ that aims to reshape the network impedances and mimic the computational weights $\mathrm{w_{k}}$ as presented:
\begin{equation}
    \mathrm{V_k^* = V_{\mathrm{ref},k}+V_{\mathrm{sec},k}+\left(R_{\mathrm{d},k}+\Delta R_{\mathrm{d},k}\right)i_k}
    \label{eq_droop_offset_appdx}
\end{equation}
where, $\mathrm{V_{\mathrm{sec},k}}$ represents the secondary offset added to the reference $V_{\mathrm{ref},k}$ to ensure consistent power flow with the rated \textit{static} case.

\subsubsection*{Targeted Computing Task}

The exemplary affine transformation task aims to rotate a 2\texttimes 2 input image by 90\textdegree, either clock-wise or counterclockwise. In this setting, a value of ``1" denotes a high-density pixel in the input image (as the shaded pixels in \textcolor{blue}{Fig.~\ref{fig_framework}}), while ``0" otherwise. The output is a scalar that represents the decimal value of the binary sequence constructed according to this pixel-to-bit assignment scheme, with top-left leading pixels corresponding to high-order digits. Mathematically, this affine transformation corresponds to an inner product $\mathrm{f(\mathbf{x})=\mathbf{w}\mathbf{x}}$ with the weights $\mathrm{[w_k] = [4, 1, 8, 2]}$ for rotating clockwise and $\mathrm{[w_k] = [2, 8, 1, 4]}$ for rotating counterclockwise.

For grid-wise implementation, we map the input image to the variation of $\mathrm{\Delta V_{ref,k}}$ by pixel in a bi-level way, such that $\mathrm{\Delta V_{ref,k}} = 1\;\text{V}$ stands for ``1" at Pixel $\mathrm{k}$ and $0\;\text{V}$ for ``0". This $\mathrm{\Delta V_{ref,k}}$ is then superimposed to the $\mathrm{V_{ref,k}}$ in (\ref{eq_droop_appdx}), namely data encoding into the grid.

\subsection*{Realization of the Computing Framework}

\begin{figure}[H]
    \centering
    \includegraphics[width=0.6\linewidth]{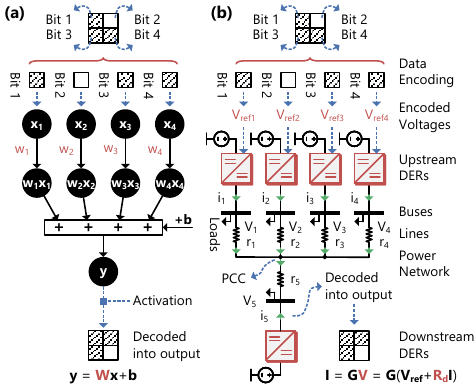}
    \caption{Exemplary implementation of an affine transformation, where (a) a $2\times2$ image is rotated (more details elaborated in \textcolor{blue}{Methods}). This can be implemented on (b) a 5-bus DC microgrid with power electronic converters (PECs) and droop-control infrastructure by superimposing virtual resistances $\mathrm{\Delta R_{d, k}}$ onto corresponding droop gain $\mathrm{R_{d, k}}$.
    In this setting, the input data are encoded as perturbations to the network (voltage reference offsets $\mathrm{\Delta V_{ref,k}}$), inducing a measurable current variation $\mathrm{\Delta i_5}$ at DER 5, which is quantized and mapped to the affine transformation result as the decoding process.
    }
    \label{fig_framework}
\end{figure}

\begin{figure}[H]
    \centering
    \includegraphics[width=0.75\linewidth]{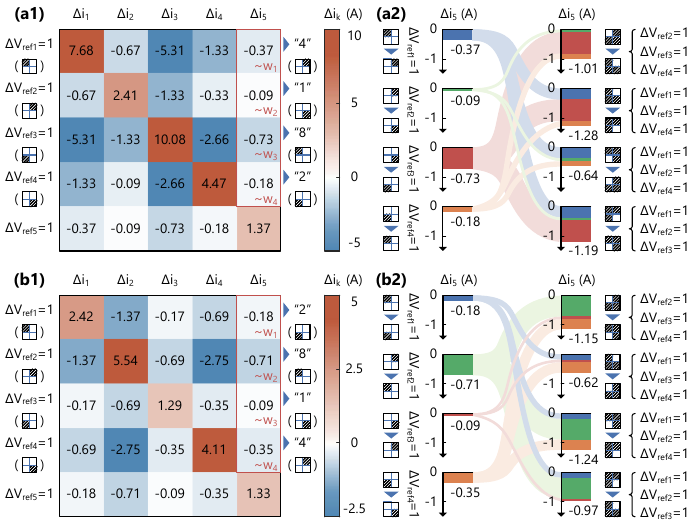}
    \caption{The source current variations $\mathrm{\Delta i_k}$ are illustrated in different input cases---\textcolor{blue}{(a1)}\&\textcolor{blue}{(a2)} for clockwise rotation, and \textcolor{blue}{(b1)}\&\textcolor{blue}{(b2)} for counterclockwise rotation, in comparison with the \textit{static} case when no data are input (i.e., $\mathrm{\Delta V_{ref,k}=0}$, $\mathrm{\forall k \in \{1,2,3,4,5\}}$). In input images, a pixel with high intensity is represented by a unit increase in corresponding $\mathrm{V_{ref,k}}$, or $\mathrm{\Delta V_{ref,k}=1\,V}$. It is indicated by one-hot input cases (\textcolor{blue}{a1}) and (\textcolor{blue}{b1}) that $\mathrm{g_{5k} = {\partial i_5}/{\partial V_{ref,k}}}$ proportionally align with the predefined computational weights $\mathrm{w_{k}}$, while (\textcolor{blue}{a2}) and (\textcolor{blue}{b2}) showcase the superposition principle with more interlaced inputs, confirming the correspondence between the intended computational transformation and the grid's physical circuit behavior.}
    \label{fig_linresults}
\end{figure}

\begin{figure}[H]
    \centering
    \includegraphics[width=0.6\linewidth]{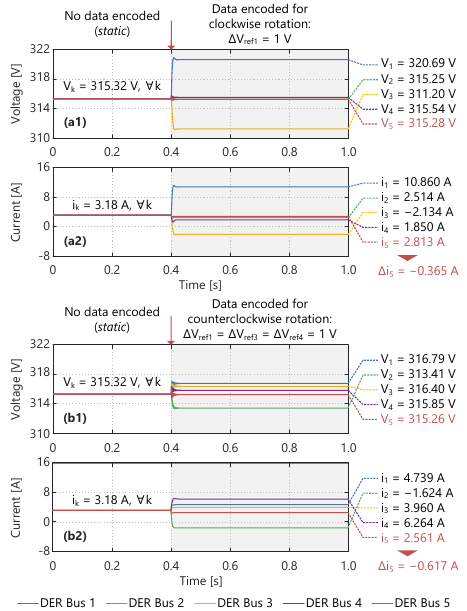}
    \caption{Realizations of affine-computation cases selected from \textcolor{blue}{Fig.~\ref{fig_linresults}} substantiating the proposed framework in the time domain. Cases are illustrated on \textcolor{blue}{(a1)}\&\textcolor{blue}{(a2)} clockwise rotation with $\mathrm{\Delta V_{ref,1}=1\,V}$, and \textcolor{blue}{(b1)}\&\textcolor{blue}{(b2)} counterclockwise rotation with $\mathrm{\Delta V_{ref,1}=\Delta V_{ref,3}=\Delta V_{ref,4}=1\,V}$, obtained from circuit-level simulations. Additional system configurations are detailed in \textcolor{blue}{Supplemental Information}.}
    \label{fig_waveforms}
\end{figure}

In a droop-based DC microgrid (droop definition and its mathematical formulation as described in \textcolor{blue}{Methods -- Case Description}), we can implement the framework accordingly as illustrated in \textcolor{blue}{Fig.~\ref{fig_framework}(b)}. Fundamentally, the terminal voltage at each DER $\mathrm{k}$ can be flexibly controlled as per the voltage-current ($\mathrm{V}$-$\mathrm{I}$) droop equation \textcolor{blue}{\cite{dcdroop}}:
\begin{equation}
    \mathrm{V_k^* = V_{\mathrm{ref},k}+R_{\mathrm{d},k}i_k},
    \quad \mathrm{k} \in \{1,2,3,4,5\}
    \label{eq_droop}
\end{equation}
where, $\mathrm{V_{\mathrm{ref},k}}$ is the rated reference at DER $\mathrm{k}$, $\mathrm{i_k}$ is the source current, and $\mathrm{V_k^*}$ is the yielded control reference for the primary voltage regulators.

The droop gains of each unit can be programmed using offsets $\mathrm{\Delta R_{\mathrm{d},k}}$, thereby shaping the network transformation characteristics (impedances). Hence, (\ref{eq_droop}) can be rewritten into:
\begin{equation}
    \mathrm{V_k^* = V_{\mathrm{ref},k}+\left(R_{\mathrm{d},k}+\Delta R_{\mathrm{d},k}\right)i_k}
    \label{eq_droop_offset}
\end{equation}

Since we are demonstrating the computational ability, we simplify the analysis by neglecting the primary control dynamics, i.e., we assume the DER terminal voltages $\mathrm{V_k}$ to track the references $\mathrm{V_k^*}$ precisely (with tuned low-level current controller):
\begin{equation}
    \mathrm{V_k^* \approx V_k}
\end{equation}

We then denote DERs 1-4 as upstream DERs responsible for input data encoding, and DER 5 as downstream DER for output data decoding. Given the DER buses coupled through a network of transmission lines, the currents flowing from DERs 1-4 to the network ($\mathrm{i_{\mathrm{in},1-4}}$) \& from the network to DER 5 ($\mathrm{i_{\mathrm{out},5}}$) (\textcolor{blue}{Fig.~\ref{fig_framework}(b)}) are derived by excluding the load currents $\mathrm{i_{\mathrm{load},k}}$ from measured source currents $\mathrm{i_k}$ as per the Kirchhoff's Current Law (KCL):
\begin{subequations}
\begin{align}
    \mathrm{i_{\mathrm{in},k}} = \mathrm{i_k-\dfrac{V_k}{R_{\mathrm{load},k}}}, &\quad \text{for upstream DERs 1-4} \\
    \mathrm{i_{\mathrm{out},5}} = \mathrm{-i_5+\dfrac{V_5}{R_{\mathrm{load},5}}}, &\quad \text{for downstream DER 5}
\end{align}
\end{subequations}

Based on the Ohm's Law, the network side satisfies:
\begin{subequations}
\begin{align}
    \mathrm{\frac{V_k-V_\mathrm{PCC}}{r_k} = i_{\mathrm{in},k}}, &\quad \text{for upstream DERs 1-4} \\
    \mathrm{\frac{V_\mathrm{PCC}-V_5}{r_5} = i_{\mathrm{out},5}}, &\quad \text{for downstream DER 5}
\end{align}
\end{subequations}
where, $\mathrm{r_k}$ represents the resistance of line $\mathrm{k}$ directly associated with corresponding upstream / downstream DERs (\textcolor{blue}{Fig.~\ref{fig_framework}(b)}), and $\mathrm{V_\mathrm{PCC}}$ is the voltage at the point of common coupling. It is worth notifying that lines 1-5 are connected together that realizes KCL from system perspective.

Using an intermediate variable $\mathrm{\lambda_k}$ for each DER $\mathrm{k}$:
\begin{equation}
    \mathrm{\lambda_k = 1-\frac{R_{\mathrm{d},k}+\Delta R_{\mathrm{d},k}}{R_{\mathrm{load},k}}},
\end{equation}
the DER terminal voltages $\mathrm{V_k}$ can be rewritten into:
\begin{subequations}
\begin{align}
    \mathrm{\mathrm{V_k} = \frac{1}{\lambda_k}\left[V_{\mathrm{ref},k}+\left(R_{\mathrm{d},k}+\Delta R_{\mathrm{d},k}\right)i_{\mathrm{in},k}\right]}, &\quad \text{for upstream DERs 1-4}\\
    \mathrm{V_5} = \mathrm{\frac{1}{\lambda_5}\left[V_{\mathrm{ref},5}-\left(R_{\mathrm{d},5}+\Delta R_{\mathrm{d},5}\right)i_{\mathrm{out},5}\right]}, &\quad \text{for downstream DER 5}
\end{align}
\end{subequations}
the current $\mathrm{i_{\mathrm{out},5}}$ flowing out of the power network at DER 5 \& equivalent conductances (including the virtual conductances as a consequence of the droop gains) are:

\begin{equation}
    \mathrm{i_{\mathrm{out},5}} = \frac{\displaystyle\sum_\mathrm{k=1}^4{\overbrace{\mathrm{\frac{1}{r_k\lambda_k-(R_{\mathrm{d},k}+\Delta R_{\mathrm{d},k})}}}^{\substack{\text{Inherent programmable}\\\text{input scaling}}}\cdot\mathrm{V_{\mathrm{ref},k}}}-\overbrace{\sum_\mathrm{k=1}^4\mathrm{\frac{\lambda_k}{r_k\lambda_k-(R_{\mathrm{d},k}+\Delta R_{\mathrm{d},k})}\cdot\frac{V_{\mathrm{ref},5}}{\lambda_5}}}^{\text{Inherent offset}}}{\underbrace{1+\sum_\mathrm{k=1}^4\mathrm{\frac{\lambda_k}{r_k\lambda_k-(R_{\mathrm{d},k}+\Delta R_{\mathrm{d},k})}\cdot\frac{r_5\lambda_5-(R_{\mathrm{d},5}+\Delta R_{\mathrm{d},5})}{\lambda_5}}}_{\text{Inherent programmable output scaling}}}
    \label{eq_iout5}
\end{equation}
\begin{equation}
    \mathrm{\frac{\partial i_{\mathrm{out},5}}{\partial V_{\mathrm{ref},k}}} = \mathrm{\frac{1}{r_k\lambda_k-(R_{\mathrm{d},k}+\Delta R_{\mathrm{d},k})}} \cdot \left[1+\sum_\mathrm{i=1}^4\mathrm{\frac{\lambda_i (r_5\lambda_5-(R_{\mathrm{d},5}+\Delta R_{\mathrm{d},5}))}{\lambda_5(r_i\lambda_i-(R_{\mathrm{d},i}+\Delta R_{\mathrm{d},i}))}}\right]^{-1}, \quad \text{for } \mathrm{k \in \{1,2,3,4\}}
    \label{eq_zeq5}
\end{equation}
and, the measured source current $\mathrm{i_5}$ at DER 5, from which the output information will be decoded, can be written as:
\begin{equation}
    \mathrm{i_5} = \mathrm{\frac{1}{\lambda_5}\left(-i_{\mathrm{out},5}+\frac{V_{\mathrm{ref},5}}{R_{\mathrm{load},5}}\right)}
\end{equation}

Throughout the derivation, it could thereby be summarized that the input data are transformed, propagated and aggregated by the physical network from DER 1-4 to DER 5 through:
\begin{equation}
    \mathbf{x} \xrightarrow{\text{Encoding}} \mathrm{\Delta V_{ref, 1-4}} \xrightarrow{\text{Line 1-4}} \mathrm{\Delta i_{in, 1-4}} \xrightarrow{\text{KCL at PCC}} \mathrm{\Delta i_{out, 5}} \xrightarrow{\text{Line 5}} \mathrm{\Delta i_{5}} \xrightarrow{\text{Decoding}} \mathbf{y}
\end{equation}
which is mathematically equivalent to the affine transformation:
\begin{equation}
    \mathbf{y} = \mathbf{w}\mathbf{x}+\mathbf{b}
\end{equation}

Therefore, the computational weights $\mathrm{w_k}$ in \textcolor{blue}{Fig.~\ref{fig_framework}} are mapped to the following proportional relationship:
\begin{equation}
    \mathrm{w_k} = \mathrm{\kappa\frac{\partial i_5}{\partial V_{\mathrm{ref},k}}} \propto \mathrm{\frac{1}{r_k\lambda_k-(R_{\mathrm{d},k}+\Delta R_{\mathrm{d},k})}}
\end{equation}
where, $\mathrm{\kappa}$ represents the common proportional coefficient for all upstream DERs 1-4. It should be properly selected such that the resulting $\mathrm{\Delta R_{\mathrm{d},k}}$ will not drive the system beyond the stability constraints delineated by the control loop dynamics.

Based on the data-image mapping scheme described in \textcolor{blue}{Methods -- Case Description}, we decode the output data from the integer multiplications of the current variation $\mathrm{\Delta i_5}$ at DER 5 (i.e., 16-level quantization for the presented 4-bit image case). Without loss of generality, by assigning $\mathrm{\Delta R_{\mathrm{d},k_0} = 0}$ for the node $\mathrm{k_0}$ such that the corresponding computational weight $\mathrm{w_{k_0} = 1}$, the remaining four $\mathrm{\Delta R_{\mathrm{d},k}}$ can be yielded as per the proportional relationship of $\mathrm{w_k}$.

Meanwhile, to preserve the consistency in \textit{static} power flow (when no input data is encoded or no pixel is activated) with the droop parameters $\mathrm{(R_{\mathrm{d},k} + \Delta R_{\mathrm{d},k})}$ obtained due to computing, an offset $\mathrm{V_{\mathrm{sec},k}}$ can be superimposed to the voltage reference $\mathrm{V_{\mathrm{ref},k}}$ as:
\begin{equation}
    \mathrm{V_k^* = V_{\mathrm{ref},k}}+ \hspace{-27pt}\underbrace{\mathrm{V_{\mathrm{sec},k}}}_{\substack{\text{Secondary offset}\\ \text{for preserving power flow}}}\hspace{-27pt}+\mathrm{\left(R_{\mathrm{d},k}+\Delta R_{\mathrm{d},k}\right)i_k}
\end{equation}
where, the offset $\mathrm{V_{\mathrm{sec},k}}$ should be equal to:
\begin{equation}
    \mathrm{V_{\mathrm{sec},k}} = \mathrm{-\Delta R_{\mathrm{d},k}i_k^0}
\end{equation}
with $\mathrm{i_k^0}$ being the output current $\mathrm{i_k}$ under \textit{static} conditions.


\newpage

\clearpage

\printbibliography

@misc{energyai,
    title={{Energy and AI}},
    author={IEA},
    year={2025},
    url={https://www.iea.org/reports/energy-and-ai},
    note={Deposited 25 Jan 2026}
}

@inproceedings{jevonparadox,
    title={{From Efficiency Gains to Rebound Effects: The Problem of Jevons' Paradox in AI's Polarized Environmental Debate}}, 
    author={Luccioni, Alexandra Sasha and Strubell, Emma and Crawford, Kate},
    booktitle={Proc. 2025 ACM Conference on Fairness, Accountability, and Transparency (FAccT'25)},
    year={2025},
    volume={},
    number={},
    pages={76–88},
    doi={10.1145/3715275.3732007}
}

@article{annualreviews,
   author={Stern, Menachem and Murugan, Arvind},
   title={{Learning Without Neurons in Physical Systems}}, 
   journal={Annu. Rev. Condens. Matter Phys.},
   year={2023},
   volume={14},
   pages={417-441},
   doi={10.1146/annurev-conmatphys-040821-113439},
   publisher={Annual Reviews}
}

@article{jaeger2023computingphys,
    author={Herbert Jaeger and Beatriz Noheda and Wilfred G. van der Wiel},
    title={{Toward a Formal Theory for Computing Machines Made out of Whatever Physics Offers}},
    journal={Nat. Commun.},
    volume={14},
    pages={4911},
    year={2023},
    doi={10.1038/s41467-023-40533-1}
}

@article{yan2024reservoir,
    title={{Emerging Opportunities and Challenges for the Future of Reservoir Computing}},
    author={Yan, Min and Huang, Can and Bienstman, Peter and Tino, Peter and Lin, Wei and Sun, Jie},
    journal={Nat. Commun.},
    volume={15},
    pages={2056},
    year={2024},
    doi={10.1038/s41467-024-45187-1}
}

@book{elishai2022NeuroEng,
    title={{Neuromorphic Engineering: The Scientist’s, Algorithms Designer’s and Computer Architect’s Perspectives on Brain-Inspired Computing}},
    author={Elishai Ezra Tsur},
    year={2022},
    publisher={CRC Press}
}

@article{benjamin2025phxresistive,
    title={{Universal Approximation Theorem for Nonlinear Resistive Networks}},
    author={Scellier, Benjamin and Mishra, Siddhartha},
    journal={Phys. Rev. Appl.},
    volume={23},
    issue={4},
    pages={044009},
    numpages={20},
    year={2025},
    month=apr,
    publisher={American Physical Society},
    doi={10.1103/PhysRevApplied.23.044009}
}

@book{kundurpowersys,
   author={Kundur, Prabha},
   title={{Power System Stability and Control}},
   publisher={McGraw-Hill},
   edition={1st},
   ISBN={978-0-07-035958-X},
   year={1994},
   type={Book}
}

@article{dirk2022revphx,
  title={{Collective nonlinear dynamics and self-organization in decentralized power grids}},
  author={Witthaut, Dirk and Hellmann, Frank and Kurths, J\"urgen and Kettemann, Stefan and Meyer-Ortmanns, Hildegard and Timme, Marc},
  journal= {Rev. Mod. Phys.},
  volume={94},
  issue={1},
  pages={015005},
  numpages={52},
  year={2022},
  month=feb,
  publisher={American Physical Society},
  doi={10.1103/RevModPhys.94.015005}
}

@article{jake2023phxspatial,
    title={{Spatial Analysis of Physical Reservoir Computers}},
    author={Love, Jake and Msiska, Robin and Mulkers, Jeroen and Bourianoff, George and Leliaert, Jonathan and Everschor-Sitte, Karin},
    journal={Phys. Rev. Appl.},
    volume={20},
    issue={4},
    pages={044057},
    numpages={9},
    year={2023},
    month=oct,
    publisher={American Physical Society},
    doi={10.1103/PhysRevApplied.20.044057}
}

@article{tcgreen2021circuitmodel,
    author={Li, Yitong and Gu, Yunjie and Zhu, Yue and Junyent-Ferré, Adrià and Xiang, Xin and Green, Timothy C.},
    journal={IEEE Trans. Power Electron.}, 
    title={{Impedance Circuit Model of Grid-Forming Inverter: Visualizing Control Algorithms as Circuit Elements}}, 
    year={2021},
    volume={36},
    number={3},
    pages={3377-3395},
    doi={10.1109/TPEL.2020.3015158}
}

@misc{entsoroadmap,
    title={{ENTSO-E Research, Development \& Innovation Roadmap 2024-2034}},
    author={ENTSO-E},
    year={2024},
    url={https://www.entsoe.eu/news/2024/07/10/entso-e-research-development-innovation-roadmap-2024-2034/},
    note={Deposited 25 Jan 2026}
}

@article{dcdroop,
    author={Sahoo, Subham and Mishra, Sukumar},
    journal={IEEE Trans. Smart Grid}, 
    title={{A Distributed Finite-Time Secondary Average Voltage Regulation and Current Sharing Controller for DC Microgrids}}, 
    year={2019},
    volume={10},
    number={1},
    pages={282-292},
    doi={10.1109/TSG.2017.2737938}
}

@article{roy2019neuromorphic,
    title={{Towards Spike-based Machine Intelligence with Neuromorphic Computing}},
    volume={575},
    ISSN={1476-4687},
    DOI={10.1038/s41586-019-1677-2},
    number={7784},
    journal={Nature},
    publisher={Springer Science and Business Media LLC},
    author={Roy, Kaushik and Jaiswal, Akhilesh and Panda, Priyadarshini},
    year={2019},
    month=nov,
    pages={607–617}
}

@article{jason2023snntraining,
    author={Eshraghian, Jason K. and Ward, Max and Neftci, Emre O. and Wang, Xinxin and Lenz, Gregor and Dwivedi, Girish and Bennamoun, Mohammed and Jeong, Doo Seok and Lu, Wei D.},
    journal={Proc. IEEE}, 
    title={{Training Spiking Neural Networks Using Lessons From Deep Learning}}, 
    year={2023},
    volume={111},
    number={9},
    pages={1016-1054},
    doi={10.1109/JPROC.2023.3308088}
}

@article{spiketalk,
    author={Sahoo, Subham},
    journal={IEEE Trans. Smart Grid}, 
    title={{Spike Talk: Genesis and Neural Coding Scheme Translations}}, 
    year={2025},
    volume={16},
    number={3},
    pages={2659-2670},
    doi={10.1109/TSG.2025.3547928}
}

@book{circuitbook,
    title={{Foundations of Analog and Digital Electronic Circuits}},
    author={Agarwal, Anant and Lang, Jeffrey H.},
    year={2005},
    publisher={Morgan Kaufmann},
    isbn={9781558607354}
}

\clearpage
\setcounter{page}{1}


\section*{Document S1. Supplemental Details}

\begin{figure}[H]
    \centering
    \renewcommand{\thefigure}{S1}
    \includegraphics[width=0.6\linewidth]{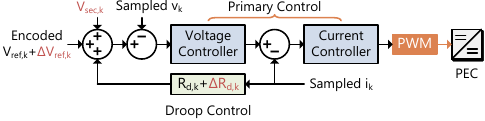}
    \caption{Block diagram of the DC droop control employed in the case study, where each droop gain $\mathrm{R_{\mathrm{d},k}}$ enabling proportional current sharing is adjusted by an offset $\mathrm{\Delta R_{\mathrm{d},k}}$ to facilitate the target affine transformation.}
    \label{fig_dcdroop}
\end{figure}
\bigskip

\subsection*{Note S1. Key parameters of the Studied Grid}

In this work, without loss of generality, we assign the key parameters of the studied electric grid as:
\begin{enumerate}[itemsep=0pt, parsep=0pt, leftmargin=*, label=(\arabic*)]
    \item Rated voltage references $\mathrm{V_{ref,k} = 315\,V}$ for all DERs;
    \item Line resistances $\mathrm{[r_k] = [0.67, 0.49, 0.83, 0.03, 0.81]\,(\Omega)}$;
    \item Load resistances $\mathrm{R_{load,k} = 99\,\Omega}$ $\mathrm{(1\,kW)}$ for all DERs;
    \item Default droop gains $\mathrm{R_{d,k} = 0.1\, V/A}$ for all DERs.
\end{enumerate}

Besides, to simplify the analysis, we assume sufficient capacity for all DERs and neglect their control dynamics. The assumptions do not compromise the validity of the proposed framework.

\bigskip

\subsection*{Note S2. Resulting Parameters from the Computing Framework}

The resulting droop gain offsets $\mathrm{\Delta R_{d,k}}$ and corresponding voltage reference offsets $\mathrm{V_{sec,k}}$ from \textcolor{blue}{Methods} are:

For clockwise rotation case:
\begin{enumerate}[itemsep=0pt, parsep=0pt, label=(\arabic*)]
    \item $\mathrm{[\Delta R_{d,k}] = [0.4688, 0, 0.6748, -0.2647, 0]\,(V/A)}$;
    \item $\mathrm{[V_{sec,k}] = [-1.4897, 0, -2.1445, 0.8412, 0]\,(V)}$.
\end{enumerate}

For counterclockwise rotation case:
\begin{enumerate}[itemsep=0pt, parsep=0pt, label=(\arabic*)]
    \item $\mathrm{[\Delta R_{d,k}] = [0.2034, 0.2969, 0, -0.2522, 0]\,(V/A)}$;
    \item $\mathrm{[V_{sec,k}] = [-0.6463, -0.9435, 0, 0.8016, 0]\,(V)}$.
\end{enumerate}

\bigskip
\subsection*{Note S3. Key Specifications in Time-Domain Simulations}

The time-domain demonstrations in \textcolor{blue}{Fig. 4} are performed by circuit-level simulation using Plexim PLECS. Apart from the parameters specified in \textcolor{blue}{Supplemental Information}, additional conditions include: (1) output voltage filters consisting of inductors and capacitors (LC filters) are installed at all DER buses $\mathrm{k}$, with $\mathrm{L_{f,k} = 2\,mH}$ and $\mathrm{C_{f,k} = 10\,\upmu F}$, $\mathrm{\forall k \in \{1,2,3,4,5\}}$; (2) proportional-integral (PI) regulators are used at all DER buses $\mathrm{k}$ for bus voltage control, with $\mathrm{K_{p,k} = 5}$, $\mathrm{\forall k \in \{1,2,3,4,5\}}$, while $\mathrm{K_{i,1} = K_{i,3} = 4000}$, $\mathrm{K_{i,2} = K_{i,4} = K_{i,5} = 1000}$.

\end{document}